\RequirePackage{fix-cm}
\pdfoutput=1
\documentclass[twocolumn, final]{svjour3}          
\smartqed  
\usepackage{graphicx}
\usepackage{latexsym}
\usepackage{amsmath}
\usepackage{hyperref}
\newcommand{\ohm}{$\mathrm{\Omega}$}
\journalname{Nonlinear Dynamics}

\begin{document}

\title{Single amplifier biquad based inductor-free Chua's circuit}


\author{Tanmoy Banerjee}


\institute{T. Banerjee \at
              Nonlinear Electronics Research Laboratory (RF and Microwave), Department of Physics, The University of Burdwan, Burdwan 713 104, West Bengal, India.\\
              \email{tanbanrs@yahoo.co.in}}

\date{Received: date / Accepted: date}

\maketitle

\begin{abstract}
The present paper reports an inductor-free realization of Chua's circuit, which is designed by suitably cascading a single amplifier biquad based active band pass filter with a Chua's diode. The system has been mathematically modeled with three-coupled first-order autonomous nonlinear differential equations. It has been shown through numerical simulations of the mathematical model and hardware experiments, that the circuit emulates the behaviors of a classical Chua's circuit, e.g., fixed point behavior, limit cycle oscillation, period doubling cascade, chaotic spiral attractors, chaotic double scrolls and boundary crisis. The occurrence of chaotic oscillation has been established through experimental power spectrum, and quantified with the dynamical measure like Lyapunov exponents. 
\end{abstract}

\section{Introduction}
\label{intro}
Design of chaotic electronic circuits offers a great challenge to the research community for the last three decades \cite{og}-\cite{ramos}. The motivation for designing a chaotic electronic circuit comes mainly from two facts: first, one can `observe' chaos, and can also control the dynamics of the circuit by simply changing the physically accessible parameters of the circuit, e.g., resistor, capacitor, voltage levels, etc.; second, there are multitude of applications of chaotic electronic oscillators starting from chaotic electronic communication to cryptography \cite{setti}-\cite{banerjee1}. A Chua's circuit is the first autonomous electronic circuit where a chaotic waveform was observed experimentally, established numerically, and proven theoretically \cite{chua}-\cite{ken3st}. Moreover, it established that chaos is not a mathematical abstraction or numerical artifact, but is  a very much realizable phenomenon.

After the advent of chaotic Chua's circuit, a large number of works have been reported on different methods of realization of this circuit. All of these realizations are mainly centered around following goals: inductor-free realization of Chua's circuit, and realization of Chua's diode. The reason behind the inductor-free realization lies in the fact that the presence of an inductor makes the circuit bulky, unsuitable for IC design, less robust, etc. In the inductor-free realization, the inductor in Chua's circuit is replaced by a general impedance converter (GIC) that requires at least two op-amps. Another approach to this end is Wien-bridge based Chua's circuit variant \cite{morgul}, where a Wien-bridge oscillator is cascaded properly with the Chua's diode. In this context, Ref. \cite{kengeneric} and \cite{kengeneric1} report a general algorithm for designing Chua's circuit, in which it has been shown that a sinusoidal oscillator can be converted into Chua's circuit by incorporating a proper type of nonlinearity. Different realizations of Chua's diode, which is the only locally active nonlinearity in the circuit, have been reported, e.g., Chua's diode using VOA \cite{kenrobust}, \cite{rocha1}, CFOA \cite{kenel}, IC realization of Chua's diode \cite{ic}, etc. Also, four element Chua's circuit has been reported in Ref.\cite{bar}, which is the minimum component Chua's circuit till date. A detailed overview on the Chua's circuit implementation can be found in Ref. \cite{for} and \cite{kil}. Recently, an inductor-free realization of the Chua's circuit based on electronic analogy has been reported \cite{rocha1}, \cite{rocha2}, which provided the advantage of exploring the system dynamics in the negative parameter space (i.e., negative values of $\alpha$ and $\beta$ of a classical Chua's circuit \cite{chua}).

Present paper reports an inductor-free realization of Chua's circuit, in which we have properly cascaded a single amplifier biquad based active band pass filter with a Chua's diode. The system has been mathematically modeled with three-coupled first-order autonomous nonlinear differential equations. It has been shown through numerical solutions of the mathematical model and real world hardware experiments that the circuit shows all the behaviors of a classical Chua's circuit, e.g., fixed point, limit cycle formation, period doubling cascade, chaotic spiral attractors, double scrolls, and boundary crisis. Occurrence of chaos has been established through Lyapunov exponents and experimental power spectrum.

The paper is organized in the following manner: next section describes the proposed circuit and its mathematical modeling. Numerical simulations, computations of nonlinear dynamical measures, e.g., Lyapunov exponents are reported in section \ref{sec:3}. Section \ref{sec:4} gives an account of the experimental results. Finally, section \ref{sec:5} concludes the outcome of the whole study.
\section{Proposed circuit and its mathematical modeling}
\label{sec:2}
\begin{figure}
  \includegraphics[width=.45\textwidth]{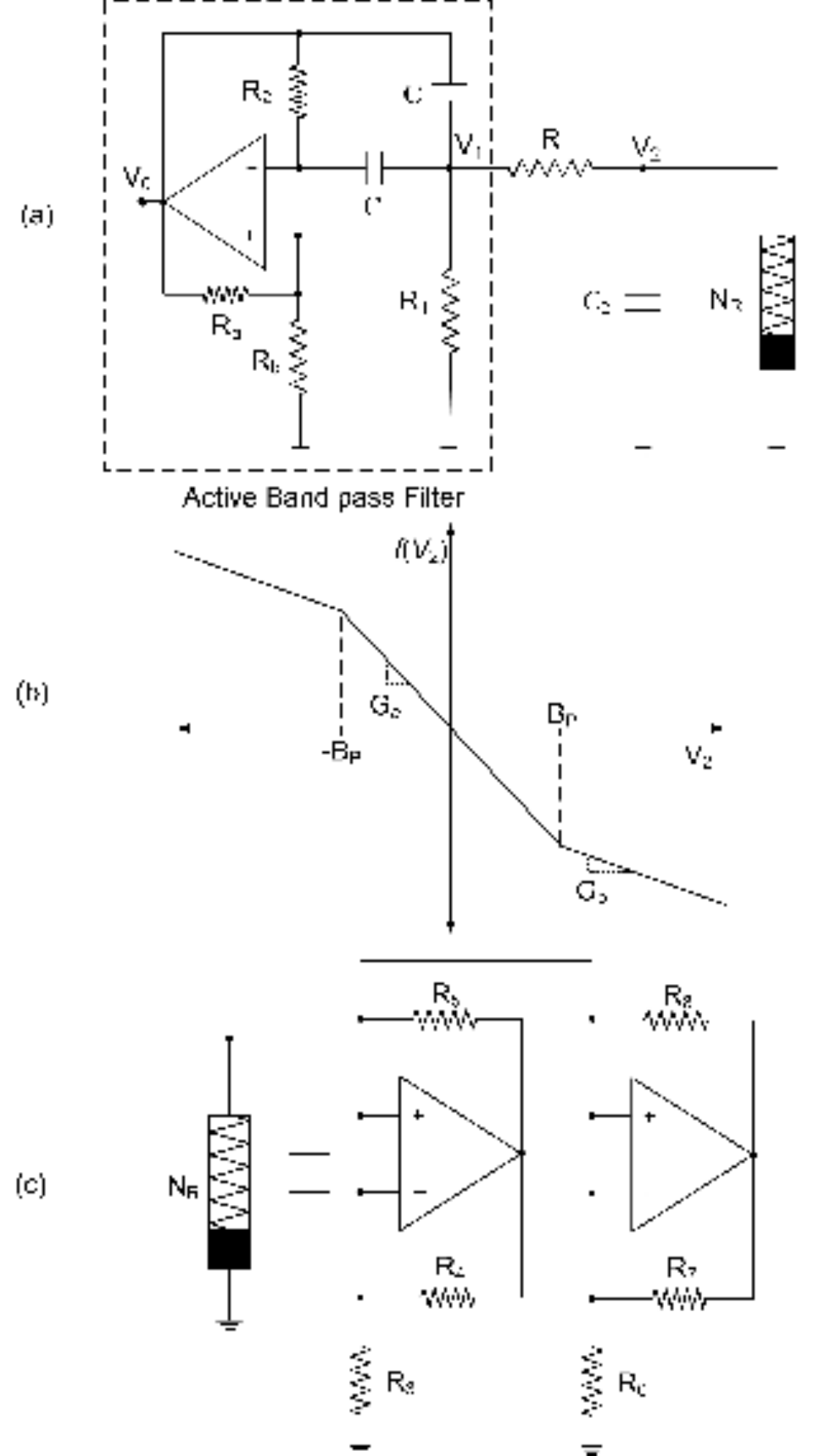}
\caption{(a) Active bandpass filter based Chua's circuit. (b) Characteristics of the Chua's diode. (c) VOA implementation of Chua's diode \cite{ken3st}, \cite{kenrobust}.}
\label{f1}       
\end{figure}

The proposed circuit is shown in Fig.\ref{f1}(a). This circuit has two distinct parts: (i) a second order narrow band active bandpass filter (BPF), and (ii) parallel combination of a grounded capacitor ($C_2$) and Chua's diode ($N_R$). $V_1$ node of the BPF is connected to the parallel combination of capacitor $C_2$ and Chua's diode ($N_R$) through a passive resistor $R$. Note that, the inductor-capacitor parallel combination of a classical Chua's circuit has been replaced by a resonator circuit, which is, in the present circuit, an active BPF; resonant (or center) frequency of the BPF can be controlled by simply varying the resistors $R_1$ or/and $R_2$ (instead of varying a capacitor, or an inductor).

To keep the number of active component lesser we have chosen the single amplifier biquad based narrow band active BPF proposed by Deliyannis and Friend \cite{del}, \cite{frnd}, which consists of only one single amplifier (in the form of an op-amp), two capacitors having same values ($C$), and four resistors ($R_1, R_2, R_a, \text{and}\, R_b$). This active BPF is a second order system with the following transfer function:
\begin{equation}
H(s)  =  \frac{-(k+1)s/R_1C}{s^2+(2/R_2C-k/R_1C)s+(1/R_1R_2C^2)}.\label{eq1}
\end{equation}
Here, $k=R_b/R_a$. Proper choice of $R_a$ and $R_b$ (and hence $k$) makes the coefficient of $s$ in the denominator negative, which in turn brings the pole of the circuit to the right half of the $s$-plane, resulting a sinusoidal oscillation. Real time dynamics of the system can be expressed in terms of the following two coupled first-order  autonomous differential equations \cite{bannd}:
\begin{subequations}
\label{eq1b}
\begin{eqnarray}
C\frac{dV_1}{dt} & = &\frac{k}{R_1}V_1-\frac{(2k+1)}{(k+1)R_2}V_0,\\
C\frac{dV_0}{dt} & = &\frac{(k+1)}{R_1}V_1-\frac{2}{R_2}V_0.
\end{eqnarray}
\end{subequations}
Equation (\ref{eq1b}) can be written as:\\
\begin{eqnarray}
\left(\begin{array}{c}
\frac{dV_{1}}{dt}\\
\frac{dV_{0}}{dt}\end{array}\right)&=&\left(\begin{array}{cc}
k/CR_{1} & -(2k+1)/(k+1)CR_{2}\\
(k+1)/CR_{1} & -2/CR_{2}\end{array}\right)\nonumber\\
&&\times\left(\begin{array}{c}
V_{1}\\
V_{0}\end{array}\right)\nonumber\\
&=&\left(\begin{array}{cc}
\alpha_{11} & \alpha_{12}\\
\alpha_{21} & \alpha_{22}\end{array}\right)\left(\begin{array}{c}
V_{1}\\
V_{0}\end{array}\right)\label{eqn1c}
\end{eqnarray}
It can be seen from (\ref{eqn1c}) that the BPF can be made to oscillate sinusoidally if one can ensure the following condition: $\alpha_{11}+\alpha_{22}=0$. This in turn gives the condition of oscillation of the sinusoidal oscillator as: $k=2R_1/R_2$; subsequently, the frequency of oscillation of the sinusoidal oscillator can be obtained from the condition:$\sqrt{\alpha_{11}\alpha_{22}-\alpha_{12}\alpha_{21}}=0$. Thus the frequency of oscillation is obtained as: $\omega_0=1/C\sqrt{R_1R_2}$.

Chua's diode is characterized by the function $f(V_2)$, where $V_2$ is the voltage drop across $C_2$. $f(V_2)$ is a three segment piece wise linear function with $G_a$ and $G_b$ as the slopes, and $B_p$ is the break point voltage for those segments. Variation of $f(V_2)$ with $V_2$ has been shown in Fig.\ref{f1}(b) (showing only the three segment region). $f(V_2)$ is defined by:
\begin{equation}\label{eq3}
f(V_2)=G_bV_2+\frac{1}{2}(G_a-G_b)(|V_2+B_p|-|V_2-B_p|).
\end{equation}
In the present work, we have chosen the VOA implementation of Chua's diode \cite{ken3st} as shown in Fig.\ref{f1}(c).

The dynamics of the proposed circuit can be described by three-coupled first-order autonomous nonlinear differential equations in terms of $V_2$, $V_1$ and $V_0$ (Fig. \ref{f1}):
\begin{subequations}
\label{eq2}
\begin{eqnarray}
C_2\frac{dV_2}{dt} & = &\frac{V_1-V_2}{R}-f(V_2),\\
C\frac{dV_1}{dt} & = &-\frac{k}{R}V_2+\frac{k}{R^{'}}V_1-\frac{(2k+1)}{(k+1)R_2}V_0,\\
C\frac{dV_0}{dt} & = &-\frac{(k+1)}{R}V_2+\frac{(k+1)}{R^{'}}V_1-\frac{2}{R_2}V_0.
\end{eqnarray}
\end{subequations}
Where, $R^{'}=R_1R/(R_1+R)$.\\
Eq.(\ref{eq2}) has been written in the following dimensionless form using dimensionless quantities: $\tau=t/RC$, $\dot{u}=\frac{du}{d\tau}$ ($u\equiv x,y,z$), $x=V_2/B_p$, $y=V_1/B_p$, $z=V_0/B_p$, $r_1=R/R_1$, $r_2=R/R_2$, $\alpha=C/C_2$: 
\begin{subequations}
\label{eq4}
\begin{eqnarray}
\dot{x}& = &\alpha[y-h(x)],\\
\dot{y}& = &-kx+k(r_1+1)y-\frac{(2k+1)r_2}{(k+1)}z,\\
\dot{z}& = &-(k+1)x+(k+1)(r_1+1)y-2r_2z.
\end{eqnarray}
\end{subequations}
Here, $h(x)$ is defined as
\begin{align}\label{eq5}
h(x)\equiv & x+f(x)\nonumber\\ 
= & m_1x+0.5(m_0-m_1)(|x+1|-|x-1|),
\end{align}
where, we have defined the following parameters: $m_0=(RG_a+1)$, and $m_1=(RG_b+1)$. 
\section{Numerical studies: Phase plane plots, Bifurcation diagrams and Lyapunov exponents}
\label{sec:3}
Numerical integration has been carried out on (\ref{eq4}) using fourth-order Runge-Kutta algorithm with step size $h=0.001$. We have chosen the resistor  $R_1$ as the control parameter (remembering that $R_1$ controls the gain and center frequency of the BPF); thus in the numerical simulations, $r_1$ acts as the control parameter keeping other parameters fixed at: $r_2=0.2$, $k=0.04$, $\alpha=20$, $m_0=-1/7$, and $m_1=2/7$. Note that, the values of $m_0$ and $m_1$ are generally used in a classical Chua's circuit.

\begin{figure}
  \includegraphics[width=.48\textwidth]{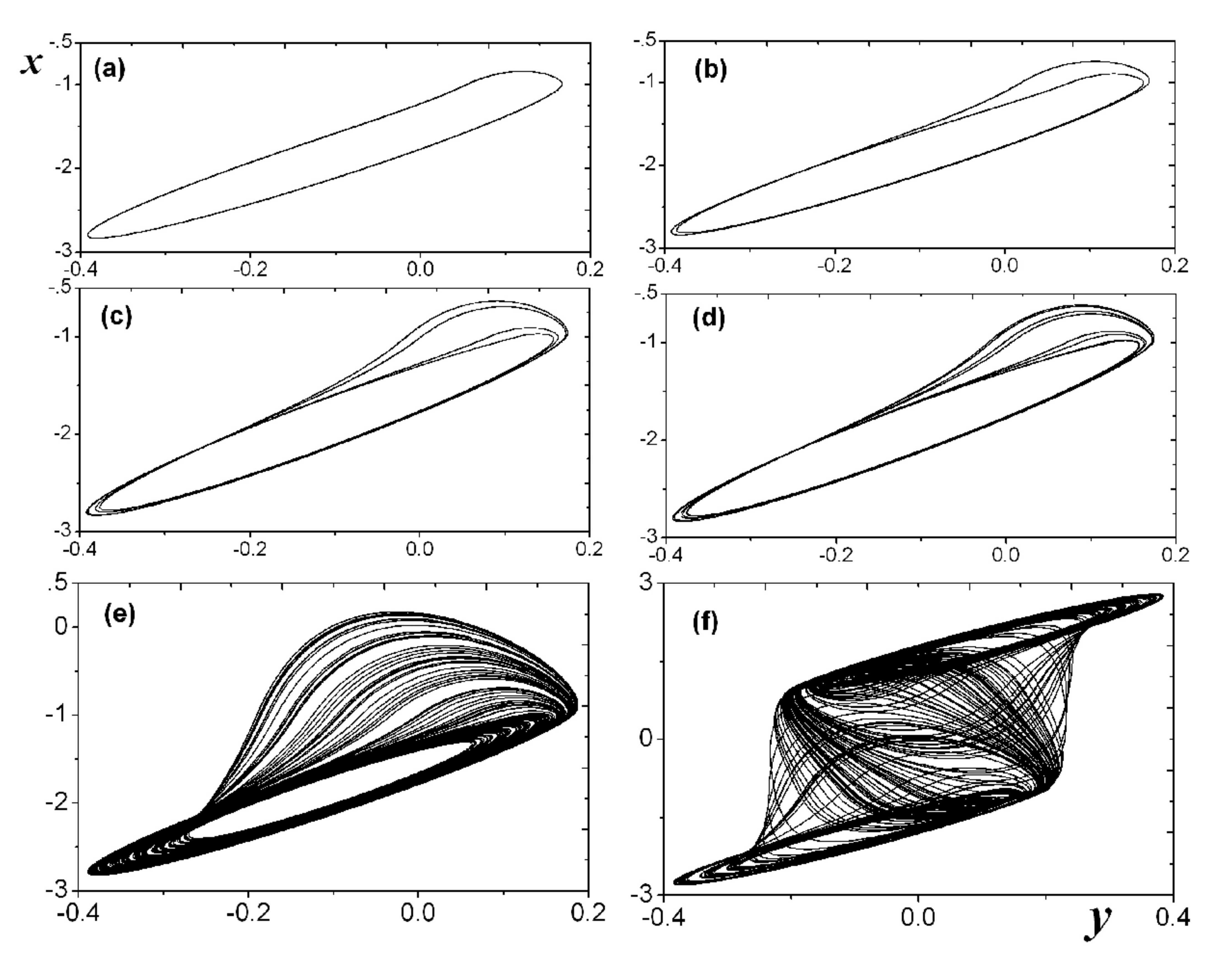}
\caption{Phase plane representation (in $y-x$ plane) for different $r_1$: (a)$r_1=15.5$ (period-1), (b)$r_1=15.6$ (period-2), (c)$r_1=15.75$ (period-4), (d) $r_1=15.77$ (period-8), (e)$r_1=16.35$(spiral chaos), (f)$r_1=17.59$ (double scroll). ($r_2=0.2$, $k=0.04$, $\alpha=20$, $m_0=-1/7$, and $m_1=2/7$)}
\label{fphyx}       
\end{figure}

\begin{figure}
  \includegraphics[width=.47\textwidth]{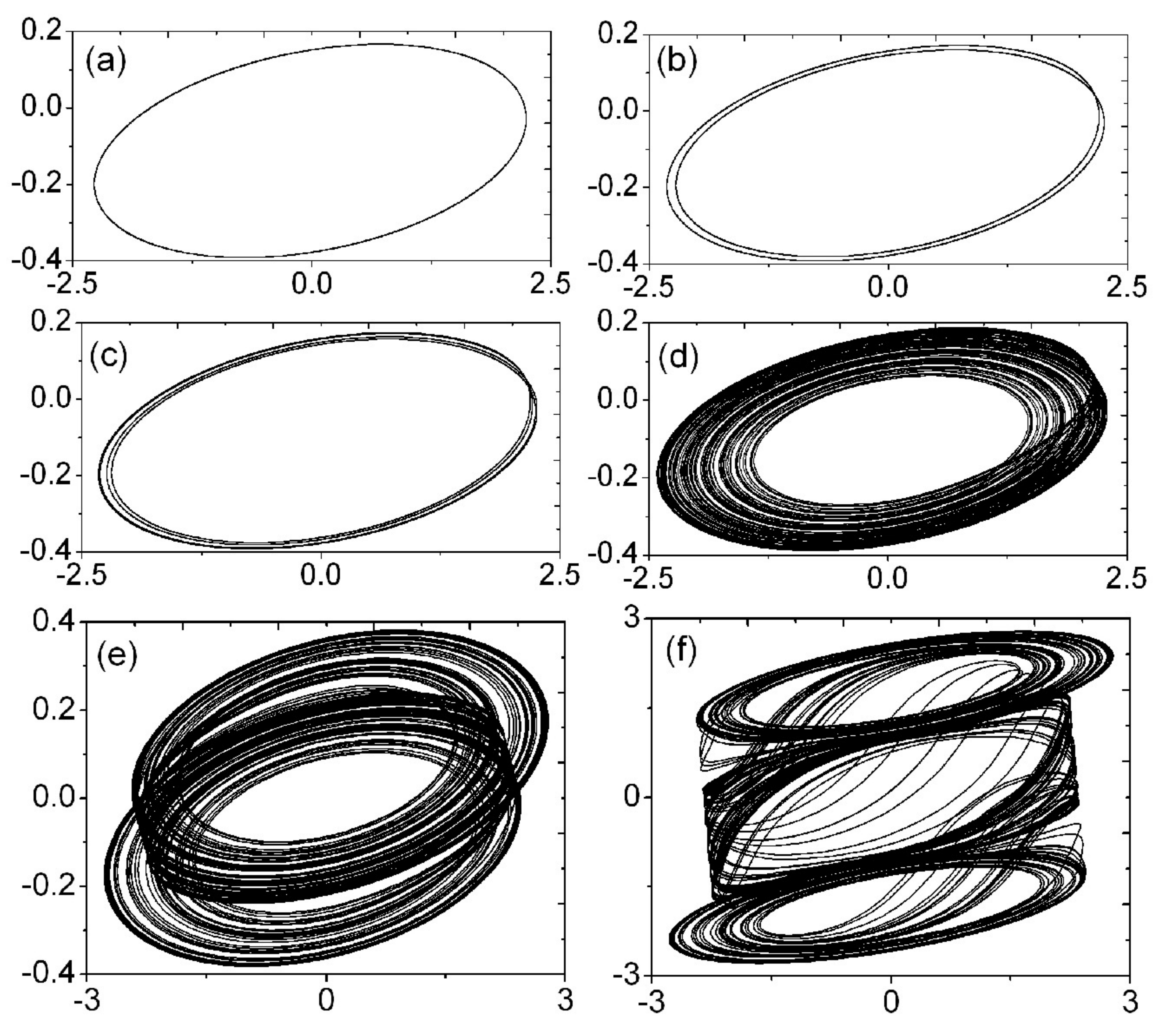}
\caption{Phase plane plots in $z-y$ plane (a)-(e), and $z-x$ plane (f) for different values of $r_1$. Parameter values are same as Fig.\ref{fphyx}.}
\label{fphzyx}       
\end{figure}

It has been observed that, with increasing $r_1$, for $r_1<15.08$, the circuit shows a fixed point. For $r_1 \ge 15.08$, the fixed point loses its stability through Hopf bifurcation, and a stable limit cycle emerges. At $r_1=15.54$, limit cycle of period-1 becomes unstable and a period-2 (P2) cycle appears. Further period doubling occurs at $r_1=15.72$ (P2 to P4), and $r_1=15.76$ (P4 to P8). Through period doubling bifurcation, the circuit enters into the regime of spiral chaos at $r_1=15.78$. With further increase of $r_1$, at $r_1=17.02$, the circuit shows the emergence of a double scroll attractor. Finally, the system-equations show diverging behavior beyond $r_1=24.81$, indicating boundary crises. Phase plane  representation (in $y-x$ plane) for different $r_1$ is shown in Fig.\ref{fphyx}, which shows the following characteristics: period-1 ($r_1=15.5$), period-2 ($r_1=15.6$), period-4 ($r_1=15.75$), period-8 ($r_1=15.77$), spiral chaos($r_1=16.35$), double scroll ($r_1=17.59$). The phase plane plots, for the same parameters, in $z-y$ plane is shown in Fig.\ref{fphzyx}(a)-(e) also, the same double scroll chaotic attractor in $z-x$ plane is shown in Fig.\ref{fphzyx}(f).

\begin{figure}
  \includegraphics[width=.45\textwidth]{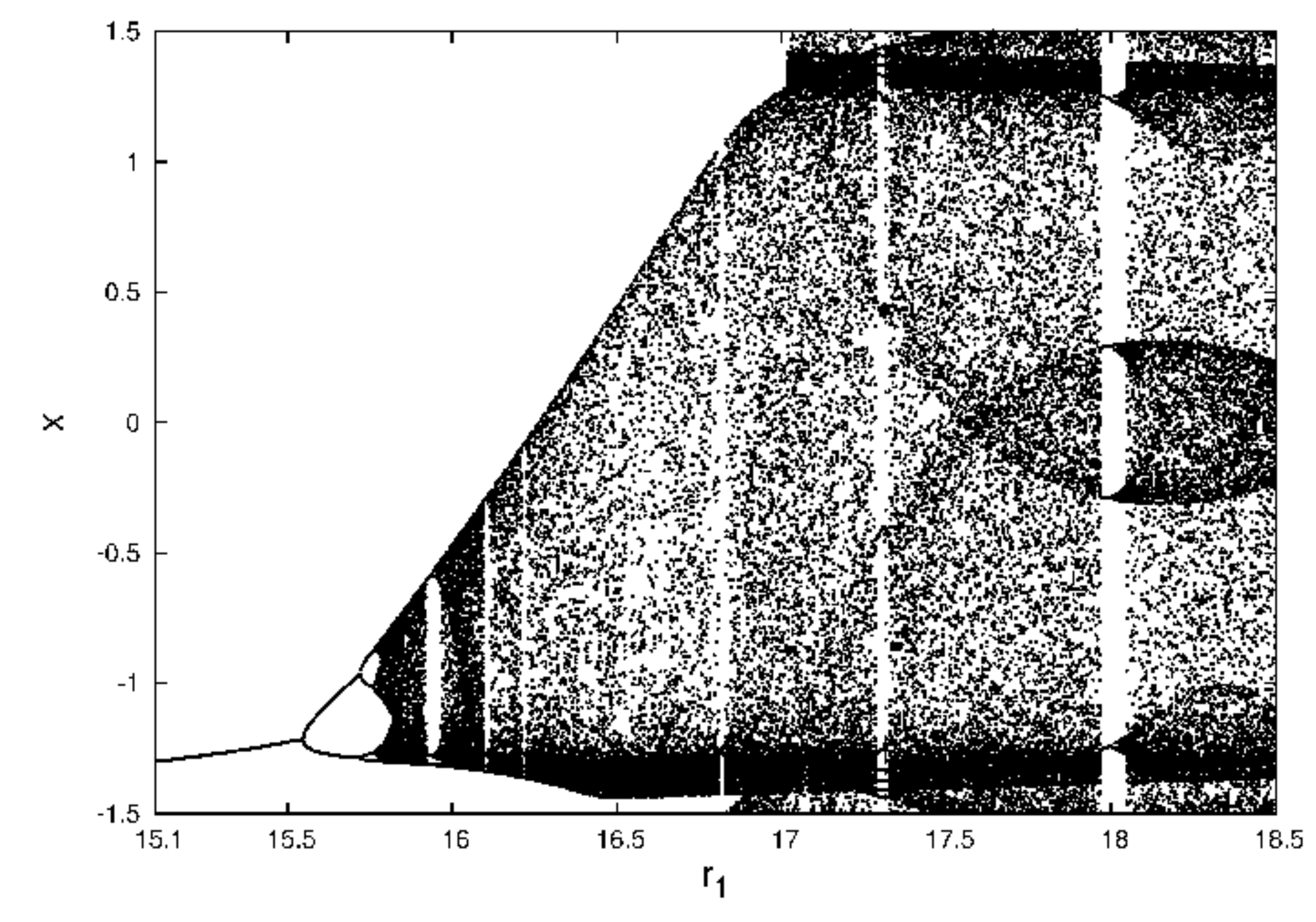}
  \includegraphics[width=.45\textwidth]{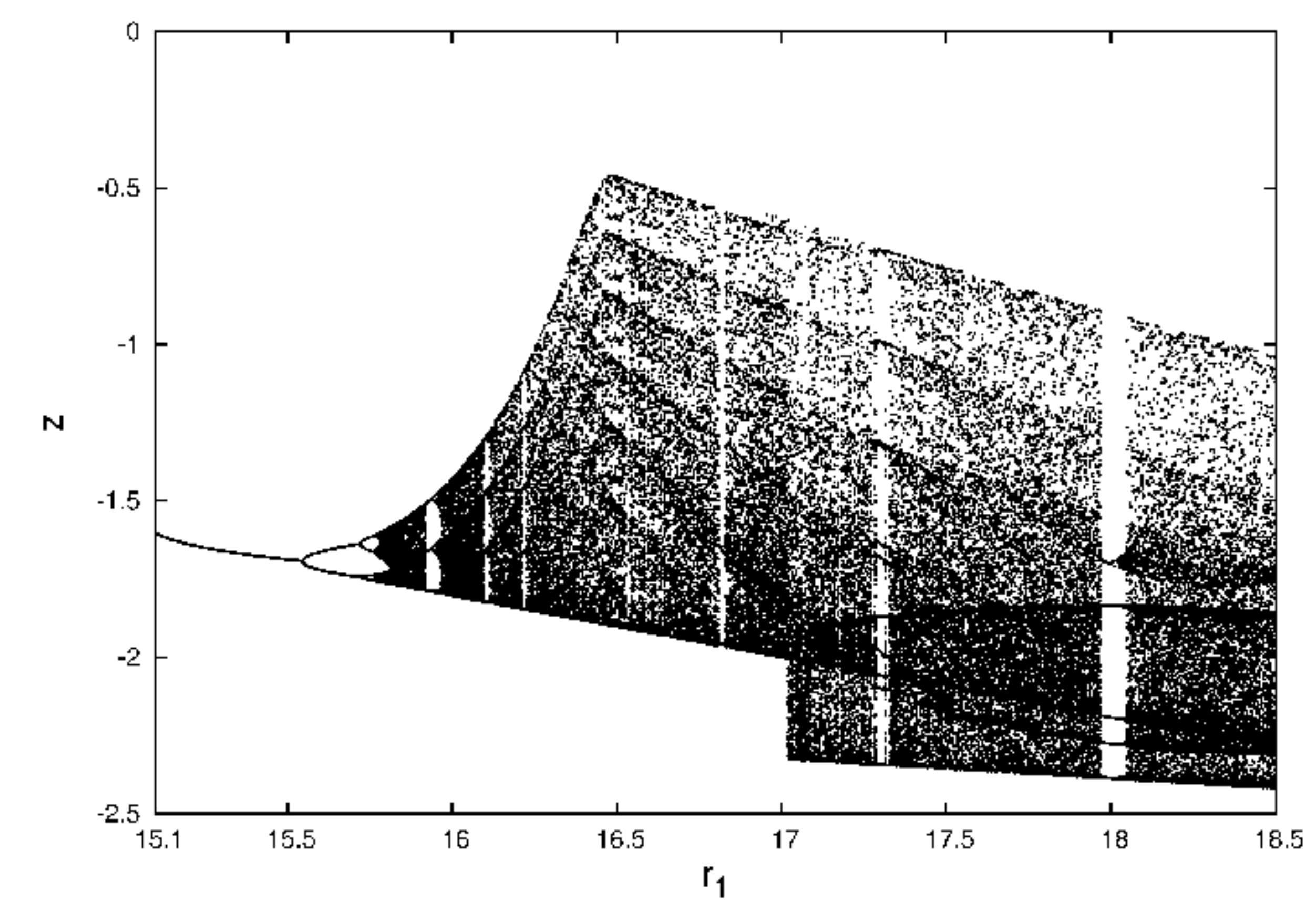}
  \includegraphics[width=.45\textwidth]{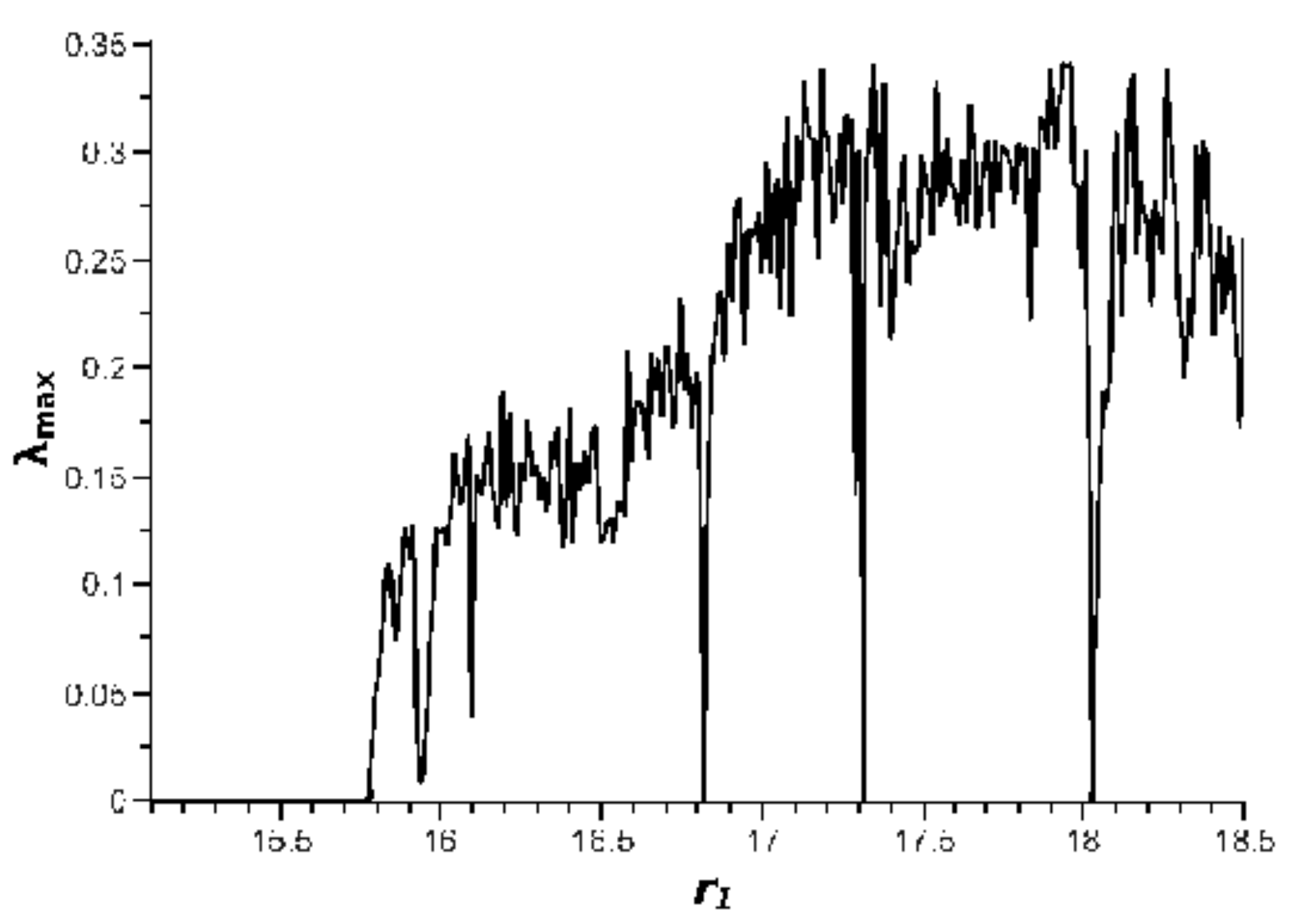}
\caption{Bifurcation diagram of $x$ (upper trace) and $z$ (middle trace) with $r_1$ as a control parameter. Lower trace shows the  Largest Lyapunov exponent ($\lambda_{max}$) with $r_1$. ($r_2=0.2$, $k=0.04$, $\alpha=20$, $m_0=-1/7$, and $m_1=2/7$).}
\label{fbif}       
\end{figure}

These observations can be summarized through bifurcation diagram with $r_1$ as the control parameter. Bifurcation diagrams are obtained by using Poincar\'{e} section \cite{ahnay} at $y=0.1$ with $\frac{dy}{dt}<0$, excluding the transients. Fig.\ref{fbif} (upper and middle trace) shows bifurcation diagrams in $x$ and $z$, respectively, for different $r_1$. Clearly, it shows a period doubling route to chaos. Further, it shows the presence of a period-3 window at $r_1=15.93$. Among the other periodic windows, more prominent is a period-5 window near $r_1=18$ interspersed in a double scroll chaos.
For quantitative measure of the chaos generated by the circuit model, Lyapunov exponents have been computed using the algorithm proposed in \cite{wolf}. Fig. \ref{fbif} (lower trace) shows the spectrum of largest Lyapunov exponent in $r_1$ parameter space. It agrees with the bifurcation diagram. Presence of a positive LE and strange attractors (as the double scroll attractor) ensures the occurrence of chaotic behavior in the circuit.

\section{Experimental results}
\label{sec:4}
The proposed circuit has been designed in hardware level on a breadboard. Since the circuit needs three VOAs we have chosen IC TL074 (quad JFET op-amp) for the present purpose with $\pm12$ volt power supply. BPF has been constructed with $C=100$ nF, $R_a=1$ k{\ohm}, $R_b=56${\ohm}, and $R_2=10$k{\ohm}. Chua's diode is constructed with following parameters \cite{ken3st}, \cite{kenrobust}: $R_3=2.2$ k\ohm,  $R_4=220$ \ohm, $R_5=220$ \ohm, $R_6=3.3$ k\ohm, $R_7=22$ k\ohm, and $R_8=22$ k\ohm. The grounded capacitor value is taken as $C_2=5$ nF. Coupling resistor $R$ is fixed at $1.3$k{\ohm} (approx.) using a $2$ k{\ohm} POT. To explore the dynamics of the circuit we have varied the resistor $R_1$ through a $1$ k{\ohm} POT. All the potentiometers are precision POT having thousand turns. Capacitors and resistors have 5\% tolerances.

\begin{figure}
  \includegraphics[width=.47\textwidth]{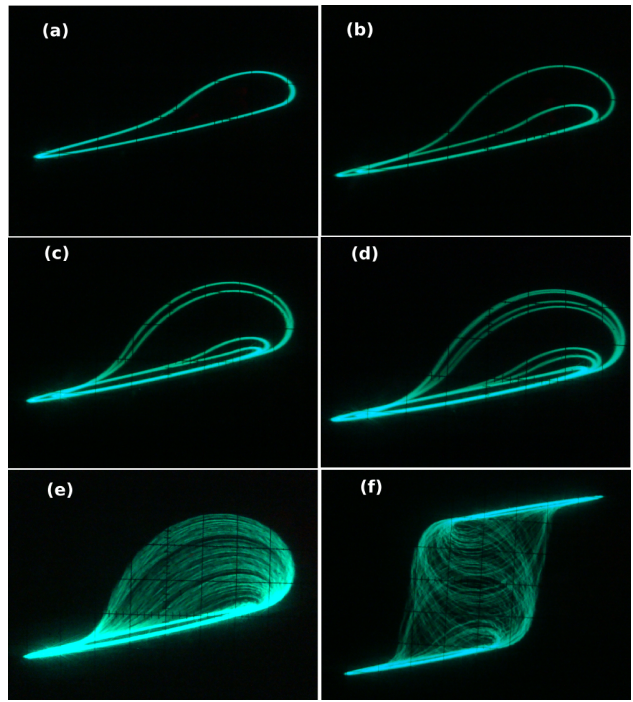}
\caption{The oscilloscope trace of experimentally obtained phase-plane plots in the $V_1$-$V_2$ space (a) Period-1 at $R_1=195${\ohm} (b) Period-2 at $R_1=180${\ohm} (c) Period-4 at $R_1=176${\ohm} (d) Period-8 at $R_1=175${\ohm} (e) Spiral chaos at $R_1=150${\ohm}. (f) Double scroll attractor at $R_1=120${\ohm}. (Other parameters are: BPF:$C=100$ nF, $R_a=1$ k{\ohm}, $R_b=56${\ohm}, and $R_2=10$k {\ohm}. Chua's diode: $R_3=2.2$ k\ohm,  $R_4=220$ \ohm, $R_5=220$ \ohm, $R_6=3.3$ k\ohm, $R_7=22$ k\ohm, and $R_8=22$ k\ohm. The grounded capacitor $C_2=5$ nF. $R=1.3$k\ohm). (a)-(e)$V_1$ ($x$ -axis): $0.1$ v/div, $V_2$ ($y$-axis): $1$ v/div. (f)$V_1$ ($x$ -axis): $0.3$ v/div, $V_2$ ($y$-axis): $2$ v/div.}
\label{fexpt1}       
\end{figure}

\begin{figure}
  \includegraphics[width=.47\textwidth]{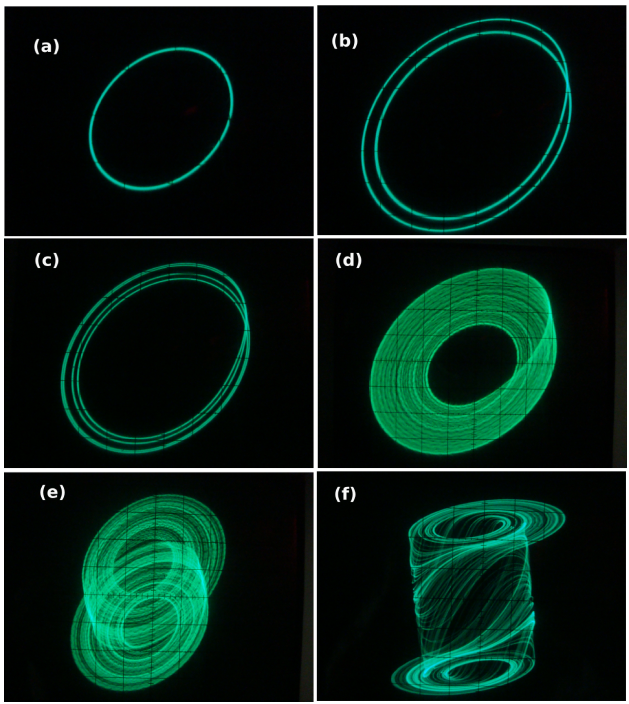}
\caption{The oscilloscope trace of experimentally obtained phase-plane plots in the (a)-(e)$V_0$-$V_1$ space ((a)-(d)$V_0$ ($x$ -axis): $1$ v/div, $V_1$ ($y$-axis): $0.1$ v/div. (e)$V_0$ ($x$ -axis): $2$ v/div, $V_1$ ($y$-axis): $0.3$ v/div.), and (f)$V_0$-$V_2$ space ($V_0$ ($x$ -axis): $2$ v/div, $V_2$ ($y$-axis): $2$ v/div.); Parameter values are same as used in Fig.\ref{fexpt1}.}
\label{fexpt2}       
\end{figure}

For $R_1>200${\ohm} the circuit shows a fixed dc value (equilibrium point of the circuit). For $R_1\le 200${\ohm} a stable limit cycle has been observed with frequency $1955$Hz. At $R_1=185${\ohm} (approx.) the limit cycle of period-1 loses its stability and a period-2 oscillation emerges. A period-4 behavior has been observed at $R_1=177${\ohm} (approx.). period-8 is found for $R_1=175${\ohm} (approx.). Further decrease in $R_1$ results in spiral chaos in the circuit (at $R_1=170$\ohm (approx.)). The double scroll attractor is observed at $R_1=122${\ohm} (approx.). The circuit shows a large limit cycle for $R_1\le68${\ohm}  that indicates the occurrence of boundary crises, where the active resistor becomes eventually passive , i.e., for a large voltage across its terminals, the $i-v$ characteristic of Chua's diode is no longer a three-segment curve but, it becomes a five-segment curve. Two additional segments (situated in two outer regions) give positive resistance behavior, i.e., now the instantaneous power consumption becomes positive \cite{ken3st}. All the above mentioned behaviors (except the large limit cycle) have been shown in Fig.\ref{fexpt1} (in $V_1$-$V_2$ space) and Fig.\ref{fexpt2} (in $V_0$-$V_1$ and $V_0$-$V_2$ space), which depict the experimental phase plane plots recorded in a real time oscilloscope (Aplab make, two channel, 60MHz).

\begin{figure}
  \includegraphics[width=.47\textwidth]{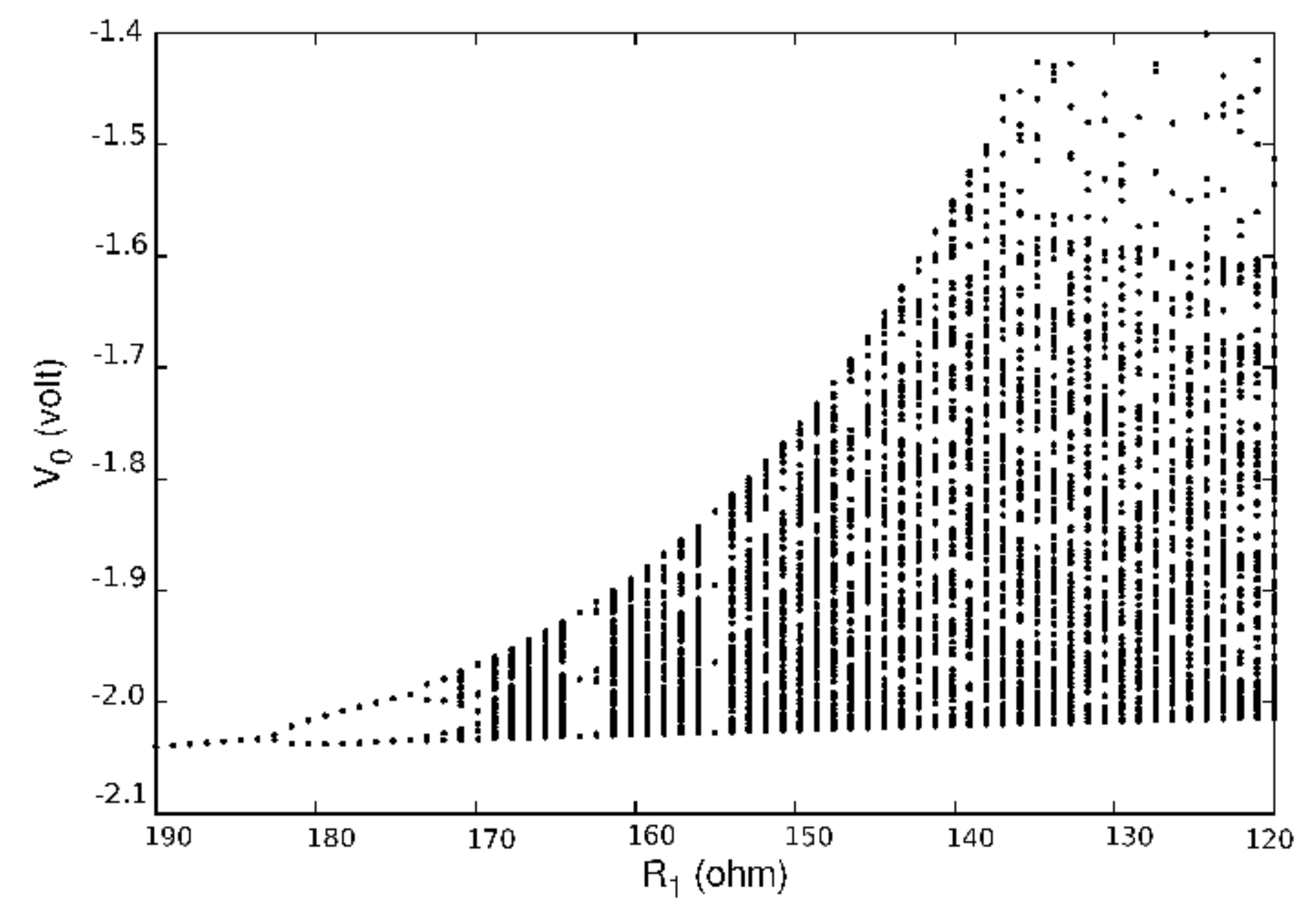}
\caption{Experimental bifurcation diagram of $V_0$ (v) with $R_1$ ({\ohm}) as the control parameter. Other parameters are same as Fig.\ref{fexpt1}.}
 \label{fexptbif}       
\end{figure}

Apart from the experimental phase plane plots, another very useful tool for exploring the real circuit behavior for a range of circuit parameters is experimental bifurcation diagram \cite{bus}, \cite{braz}. To grab the behavior of the proposed circuit for a range of values of $R_1$, experimental bifurcation diagram  of $V_0$ has been plotted taking $R_1$ as the control parameter (Fig.\ref{fexptbif}). For seventy different values of $R_1$ in the range of $195$ {\ohm} to $120$ {\ohm}, we have acquired the experimental time series data of $V_0$ using Agilent make Infinium digital storage oscilloscope (5000 data per set) and find out the poincar\'{e} section using the {\it local minima} of the time series data. Plotting these for all the $R_1$s gives the experimental bifurcation diagram. Schematic difference between the numerical bifurcation diagram (Fig.\ref{fbif} (middle trace)) and experimental bifurcation diagram  occurs due to the difference in the sampling schemes employed in finding out the Poincar\'{e} section. However, qualitatively they agree with each other, e.g., both the diagrams show period doubling route to chaos. Blurring of points in the experimental bifurcation diagram occurs due to the inherent circuit noise that is reflected in the experimental time series data. Also, since we have to acquire the data manually, resolution of the plot is not high enough in comparison with the numerical bifurcation plot.

\begin{figure}
  \includegraphics[width=.45\textwidth]{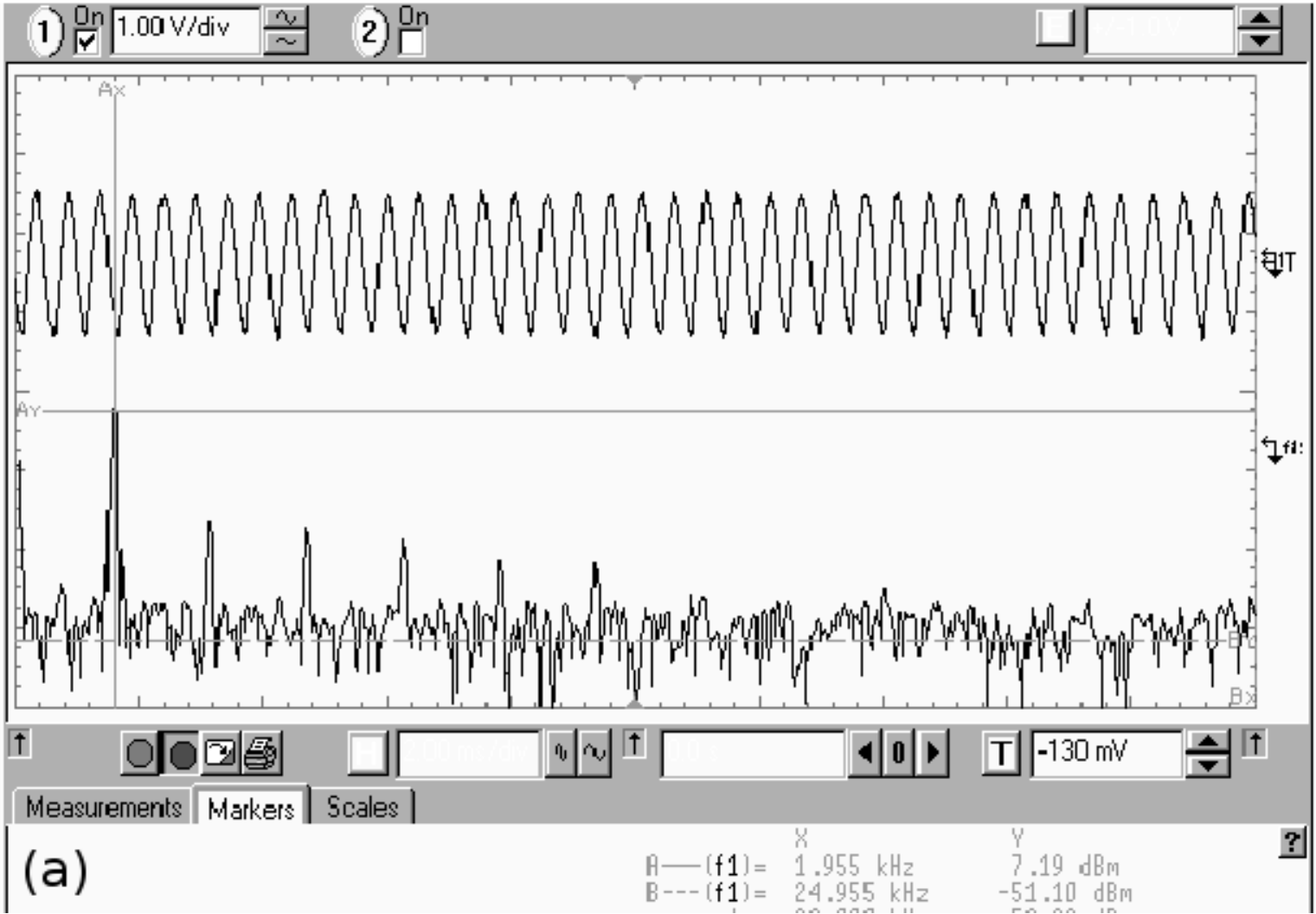}
\includegraphics[width=.45\textwidth]{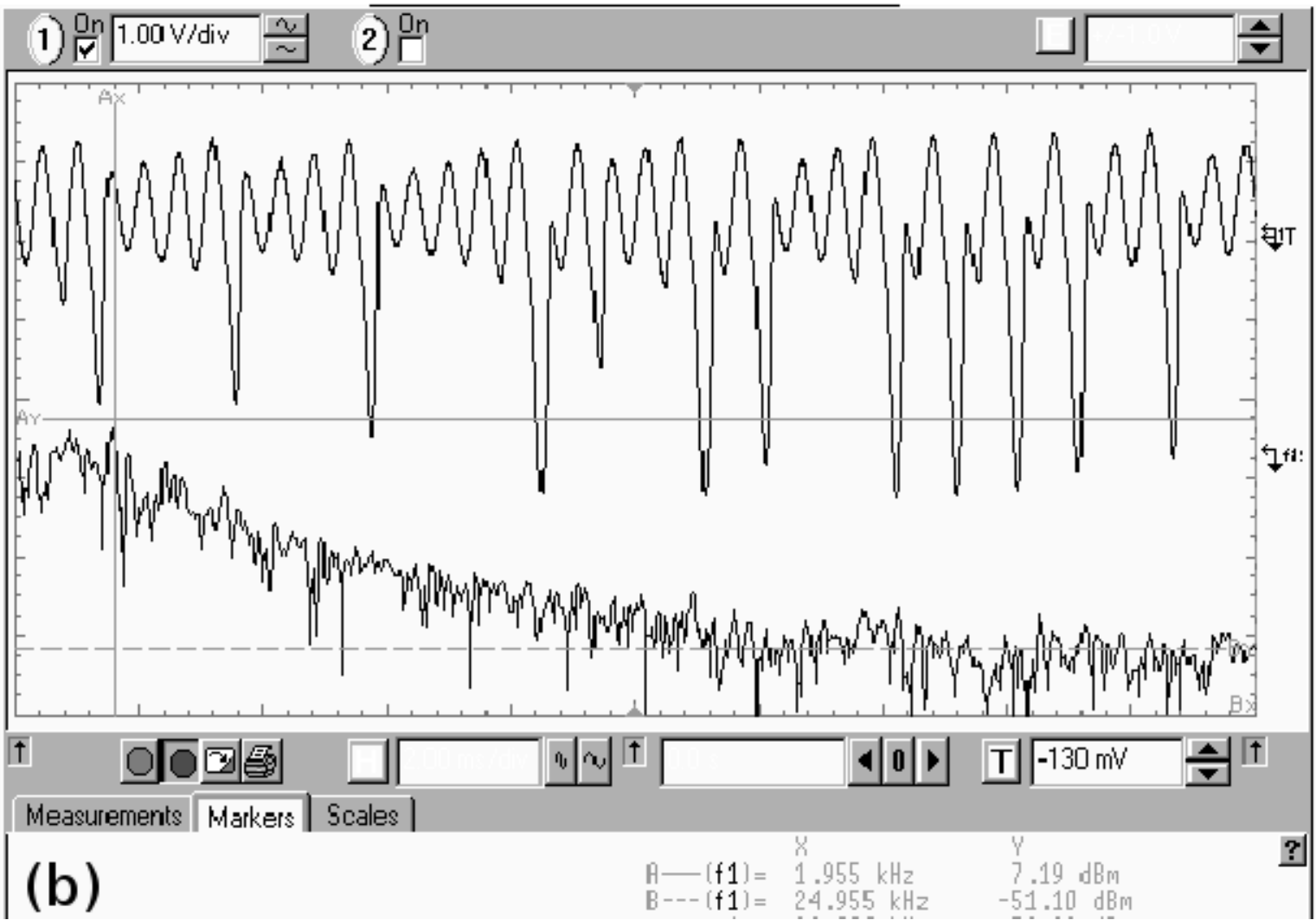}
\includegraphics[width=.45\textwidth]{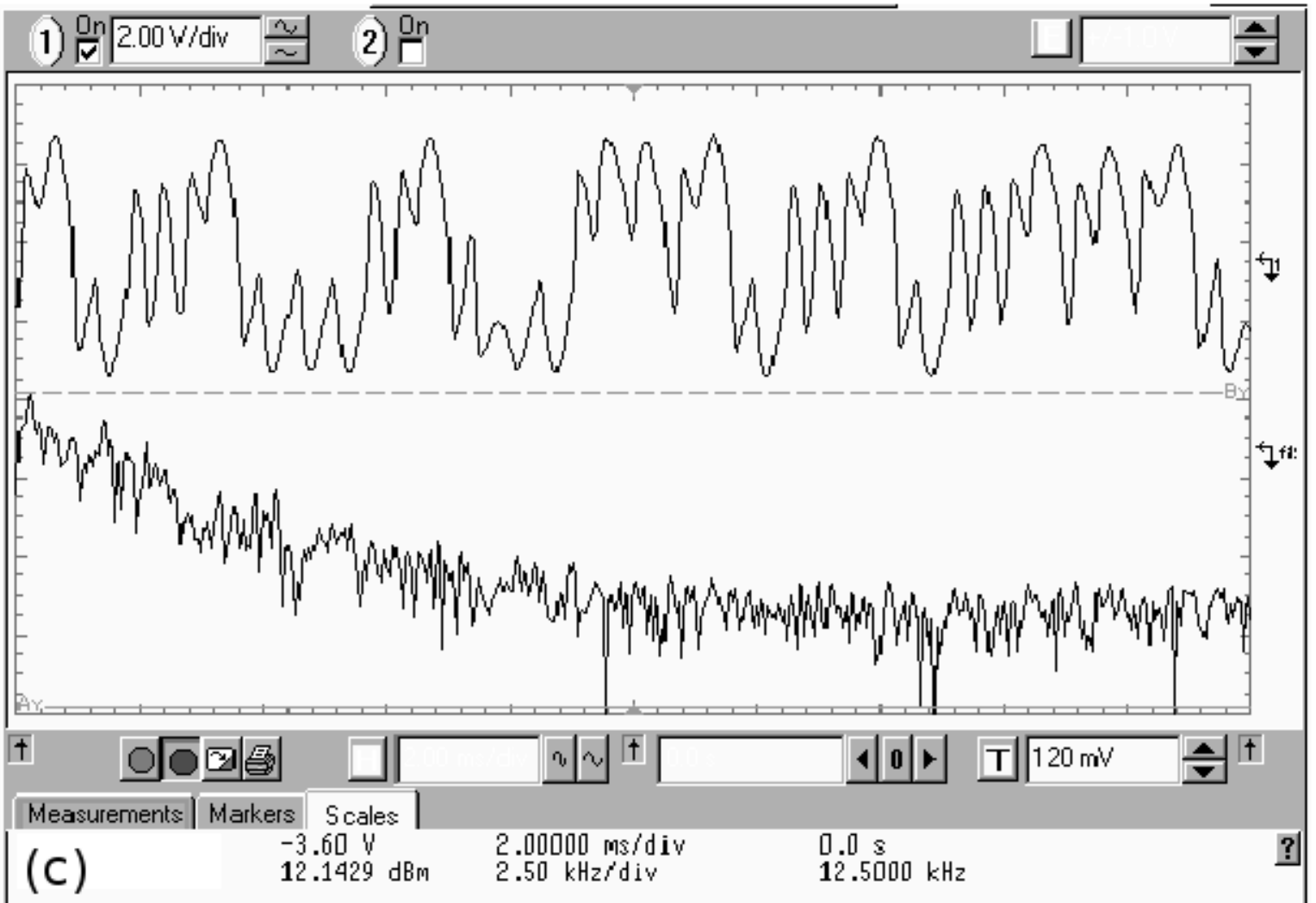}
\caption{Experimental time series and FFT of $V_2$ for (a) Period-1 oscillation (frequency $1955$Hz) (b)  spiral chaotic oscillation  and (c) double scroll chaos  (circuit parameters are same as described in Fig.\ref{fexpt1}). (Frequency span of the FFT: $25$ kHz).}
\label{fex3}       
\end{figure}

Time series data and the corresponding FFTs of $V_2$ are measured using Agilent make Infinium digital storage oscilloscope (500MHz, sampling rate 1Gs/s) and are shown in Fig.\ref{fex3}. Fig.\ref{fex3}(a) (upper trace is for time series and the lower trace is for FFT) shows the period-1 oscillation  with frequency $1955$Hz; Fig.\ref{fex3}(b) and (c) shows the results for spiral chaos and double scroll chaos, respectively. Power spectrum of Fig.\ref{fex3}(b) and (c) (lower traces) are continuous and broad in nature, indicating the occurrence of chaotic oscillations. It can be seen that the experimental results agree with numerical simulations of the mathematical model of the circuit.


\section{Conclusion}
\label{sec:5}
In this paper we have reported a single amplifier biquad based inductor-free implementation of Chua's circuit, where we have suitably cascaded an active band pass filter with a Chua's diode. The proposed circuit has been modeled mathematically by three-coupled first-order autonomous nonlinear differential equations. Numerical simulations of the mathematical model of the circuit and the real world hardware experiments confirm that the proposed circuit shows the same behavior that is shown by a classical Chua's circuit (e.g., fixed point behavior, limit cycle formation through Hopf bifurcation, period doubling, spiral chaotic attractor, double scroll attractor, and boundary crisis). The presence of chaos has been established through the Lyapunov exponents  and experimental power spectrum.

As the circuit is inductor-free it has got all the advantages of an inductor-free circuit (e.g. suitable for IC design, robustness, etc.). Further, instead of varying an inductor or a capacitor, we have used a single resistor (e.g., $R_1$) as a control parameter to observe all the complex behaviors of the circuit. Since the circuit has a large number of choices of control parameters (viz. $R_2$, $R$ and $\alpha$), one can use these parameters to observe different behaviors of the circuit. But, the basic bifurcation routes to chaos remains the same for each of the parameters. The BPF circuit we have used is narrow band (unlike Wien-bridge based circuit\cite{morgul}, where the frequency selective network is not narrow band), thus one has more control over the center frequency of the circuit. The present circuit can be suitably designed (with high frequency op-amps) to generate chaotic oscillations in a high frequency region, and also may be useful for chaos based communication systems.

\begin{acknowledgements}
The author is indebted to Prof. B.C. Sarkar (Dept. of Physics, The University of Burdwan, India) for the useful suggestions and insightful discussions. Also, the author would like to thank the anonymous reviewers for their useful suggestions.
\end{acknowledgements}



\begin{thebibliography}{99}
\bibitem{og} Ogorzalek, M.J.:Chaos and  Complexity in nonlinear electronic circuits. World Scientific Series on Nonlinear Science, Series A - Vol. 22 (1997)
\bibitem{ramos} Ramos, J.S.: Introduction to nonlinear dynamics of electronic systems: tutorial. Nonlinear Dyn. 44. 3–14 (2006)
\bibitem{setti}Kennedy, M.P., Rovatti, R. and Setti, G. (eds.), Chaotic Electronics in Telecommunications. Florida: CRC Press. (2000)
\bibitem{banerjee1}Banerjee, T., Sarkar, B.C.: Chaos, intermittency and control of bifurcation in a ZC2-DPLL. Int. J. Electron. 96(7), 717-731 (2009)
\bibitem{chua}Matsumoto, T., Chua, L.O., Komuro, M.: The double scroll.IEEE Trans. Circuits Syst. 32, 797-818 (1985)
\bibitem{ken3st} Kennedy, M.P.:Three steps to chaos-Part II:A Chua's circuit primer. IEEE Trans. Ckt. and Syst-I. 40(10), 640-656 (1993)
\bibitem{morgul}Morgul, O.: Inductorless realization of Chua's oscillator. Electron. Lett. 31, 1424-1430 (1995)
\bibitem{rocha1}Rocha, R., Medrano-T., R.O.:An inductor-free realization of the Chua's circuit based on electronic analogy. Nonlinear Dyn. 56(4), 389-400 (2009)
\bibitem{kengeneric}Elwakil, A.S, Kennedy, M.P.:Chua's circuit decomposition: a systematic design approach for chaotic oscillators. Journal of the Franklin Institute. 337, 251-265 (2000)
\bibitem{kengeneric1}Elwakil, A.S, Kennedy, M.P.:Generic RC realizations of Chua's circuit. Int. J. Bifurcation Chaos. 10, 1981-1985 (2000)
\bibitem{kenrobust}Kennedy, M.P.: Robust op-amp realization of Chua's circuit. Frequenz 46, 66-80 (1992)
\bibitem{kenel}Elwakil, A.S., Kennedy, M.P.: Improved implementation of Chua's chaotic oscillator using current feedback op-amp. IEEE Trans. Circuits Syst. I 47, 289-306 (2000)
\bibitem{ic}Cruz, J.M., Chua, L.O.: A CMOS IC nonlinear resistor for Chua's circuit. IEEE Trans. Circuits Syst. I 39, 985-995 (1992)
\bibitem{bar}Barboza, R., Chua, L. O.:The four-element Chua's circuit. Int. J. Bifurcation and Chaos 18, 943-955 (2008)
\bibitem{for}Fortuna, L., Frasca, M., Xibilia, M.G.: Chua's circuit implementations: yesterday, today, and tomorrow. World Scientific Series on Nonlinear Science Series A - Vol. 65, (2009)
\bibitem{kil}Kilic, R.: A comparative study on realization of Chua's circuit: hybrid realizations of Chua's circuit combining the circuit topologies proposed for Chua’s diode and inductor elements. Int. J. Bifurcation and Chaos 13, 1475-1493 (2003)
\bibitem{rocha2}Rocha, R., Andrucioli, G.L.D.,Medrano-T., R.O.: Experimental characterization of nonlinear systems: a real-time evaluation of the analogous Chua's circuit behavior. Nonlinear Dyn. 62(1-2),237-251 (2010)
\bibitem{del}Deliyannis, T.: High-Q factor circuit with reduced sensitivity. Electron. Lett. 4(26), 577-678 (1968)
\bibitem{frnd}Friend, J.J.: A single operational-amplifier biquadratic filter section. In: IEEE Int. Symp. Circuit Theory, pp. 189-190 (1970)
\bibitem{bannd}Banerjee, T., Karmakar, B., Sarkar, B.C.:Single amplifier biquad based autonomous electronic oscillators for chaos generation. Nonlinear Dyn. 62, 859-866 (2010)
\bibitem{ahnay}Nayfeh, A.H., Balachandran, B.: Applied Nonlinear Dynamics: Analytical, Computational, and Experimental Methods. Wiley, New York (1995)
\bibitem{wolf}Wolf, A., Swift, J. B., Swinney, H. L. and Vastano, J. A.: Determining Lyapunov Exponents from a Time Series. Physica D. 16, 285-317 (1985)
\bibitem{bus} Buscarino, A., Fortuna, L., Frasca, M. and Sciuto, G.:Coupled Inductors-Based Chaotic Colpitts Oscillator. Int. J. Bifurcation and Chaos. 2, 569-574 (2011)
\bibitem{braz} Viana, E.R., Rubinger, R.M., Albuquerque, H.A., de Oliveira, A.G. and Ribeiro, G.M.:High resolution parameter space of an experimental chaotic circuit. Chaos 20, 023110 (2010) 

\end{thebibliography}
\end{document}